\documentclass[10pt, conference, compsocconf]{IEEEtran}
\usepackage{graphicx}
\usepackage{array}
\usepackage{multirow}
\usepackage{amsmath}
\usepackage{algorithm}
\usepackage{algorithmic}
\usepackage{subfigure}
\usepackage{float}
\begin{document}

\title{Minimizing Electricity Theft\\ using Smart Meters in AMI}

\author{M. Anas, N. Javaid, A. Mahmood, S. M. Raza, U. Qasim$^{\ddag}$, Z. A. Khan$^{\S}$\\

        $^{\ddag}$University of Alberta, Alberta, Canada\\
        Department of Electrical Engineering, COMSATS\\ Institute of
        Information Technology, Islamabad, Pakistan. \\
        $^{\S}$Faculty of Engineering, Dalhousie University, Halifax, Canada.
        }

\maketitle

\begin{abstract}
Global energy crises are increasing every moment. Every one has the attention towards more and more energy production and also trying to save it. Electricity can be produced through many ways which is then synchronized on a main grid for usage. The main issue for which we have written this survey paper is losses in electrical system. Weather these losses are technical or non-technical. Technical losses can be calculated easily, as we discussed in section of mathematical modeling that how to calculate technical losses. Where as non-technical losses can be evaluated if technical losses are known. Theft in electricity produce non-technical losses. To reduce or control theft one can save his economic resources. Smart meter can be the best option to minimize electricity theft, because of its high security, best efficiency, and excellent resistance towards many of theft ideas in electromechanical meters. So in this paper we have mostly concentrated on theft issues.
\end{abstract}


\section{Introduction}
\PARstart{E}{lectricity}, generated through many ways, is synchronized on a single bus bar of the grid for transmission. Before utilization of electricity, it passes from certain phases. It is first generated, step upped in transformer deck, passed from switch yard for transmission through power lines. After transmission it is distributed for utilization to the customers. This energy needs to be billed as well. Usually two types of devices are mainly used for billing procedure.

\begin{enumerate}
\item
Electromechanical KWh meters.
\item
Smart meters.
\end{enumerate}

Our energy is strained to the utmost now a day, so using energy efficiently is one of the issues which need urgent attention. That is why electricity is to be dealt with great care. As for as knowledge is concerned there is no such password which can not be cracked but best password is the one which is being cracked in a larger period of time. This is one basic reason that whole world is shifting from analog devices to digital devices. That is why analog electromechanical meters are being substituted by smart meters.
Digital devices provide better security and controlling options. The better detection and controlling of losses is one of the reasons for substitution of smart meters.

Every thing occurs for a reason, so the reason for this substitution is losses in electrical systems. There are mainly two types of losses.

\begin{enumerate}
\item
Technical losses.
\item
Non-Technical/Commercial losses.
\end{enumerate}

In developing countries electricity theft is a common practice specially in remote areas, as they do not pay utility bills to a government company in case of electricity and gas as well. To solve this problem governments must think of an idea to provide help in terms of subsidy to manage this issue.

In section-II related work and motivation is explained. In section-III of this paper losses are discussed which are caused due to electricity theft. In section-IV ways of communication are discussed, to send data from end user to the grid. In section-V causes and effects of electricity theft is explained. Some mathematical techniques are discussed in section VI. In section VII we have concluded this paper.

\section{Related Work and Motivation}
In [1,3] authors explained theft control very well in a sense that they proposed a model. In this model they calculated NTL in external control section, and if NTL \textgreater\ 5\%, legal customers are disconnected for some interval. Harmonic generator is operated in this time period, which destroys the electrical equipment of all the illegal consumers. Reconnect normal supply for genuine customers. Although this is a good model that electricity theft is an issue that one can make equipments of an illegal users starts malfunctioning. However this model can be improved to stop functioning of the equipment of an illegal users, weather using smart meters or any other technique.

S. McLaughlin \textit{et al.} explained some of the energy theft in Advanced Metering Infrastructure (AMI), proposed an idea of a communication architecture from smart meter to grid using meter to meter communication. For boosting the data signals using collectors and receptors. It defines this procedure in a network known as Backhaul network, used to transport data to utility. However energy theft in smart meters can be a technical person, if he removes the $\mu$-controller from his meter. It will not be able to measure readings and send it to utility for further process.

\begin{table*}[t]
\begin{center}
\begin{tabular}{|m{2cm}|c|c|r|r|r|r|r|r|r|r|}
\hline
\multicolumn{ 3}{|c|}{{\bf Months }} & {\bf July} & {\bf Aug} & {\bf Sept} & {\bf Oct} & {\bf Nov} & {\bf Dec} & {\bf Jan} & {\bf Feb} \\
\hline
\multicolumn{ 1}{|c|}{{\bf 2010-2011}} & \multicolumn{ 1}{|c|}{{\bf Energy (MKWH)}} & {\bf Received} &  1764.81 &  1777.29 &  1518.89 &  1461.89 &  1136.25 &  1179.97 &  1169.85 &  1058.03 \\ \cline{3-11}

\multicolumn{ 1}{|c|}{{\bf }} & \multicolumn{ 1}{|c|}{{\bf }} & {\bf Sold} &  1508.41 &  1513.76 &  1311.82 &  1282.98 &  1047.91 &  1060.11 &  1057.74 &  1009.38 \\
\cline{2-11}
\multicolumn{ 1}{|c|}{{\bf }} & \multicolumn{ 2}{|c|}{{\bf Percentage Losses}} &    14.53 &    14.83 &    13.63 &    12.24 &     7.77 &    10.16 &     9.58 &     4.60 \\
\hline
\multicolumn{ 1}{|c|}{{\bf 2011-2012}} & \multicolumn{ 1}{|c|}{{\bf Energy (MKWH)}} & {\bf Received} &  1693.09 &  1768.82 &  1570.68 &  1509.01 &  1199.71 &  1179.12 &  1127.43 &  1140.52 \\ \cline{3-11}

\multicolumn{ 1}{|c|}{{\bf }} & \multicolumn{ 1}{|c|}{{\bf }} & {\bf Sold} &  1449.12 &  1510.51 &  1365.99 &  1329.63 &  1106.89 &  1115.94 &  1024.03 &  1085.20 \\
\cline{2-11}
\multicolumn{ 1}{|c|}{{\bf }} & \multicolumn{ 2}{|c|}{{\bf Percentage Losses}} &    14.41 &    14.60 &    13.03 &    11.89 &     7.74 &     5.36 &     9.17 &     4.85 \\
\hline
\multicolumn{ 3}{|c|}{{\bf Decrease}} &     0.12 &     0.22 &     0.60 &     0.35 &     0.04 &     4.80 &     0.41 &    -0.25 \\
\hline
                                  \multicolumn{ 11}{c}{Table 1. Energy Losses in a populated city in year 2010 till 2012} \\
\end{tabular}
\end{center}
\end{table*}

\begin{figure*}[t]
\centering
\includegraphics[height=10cm, width=14cm]{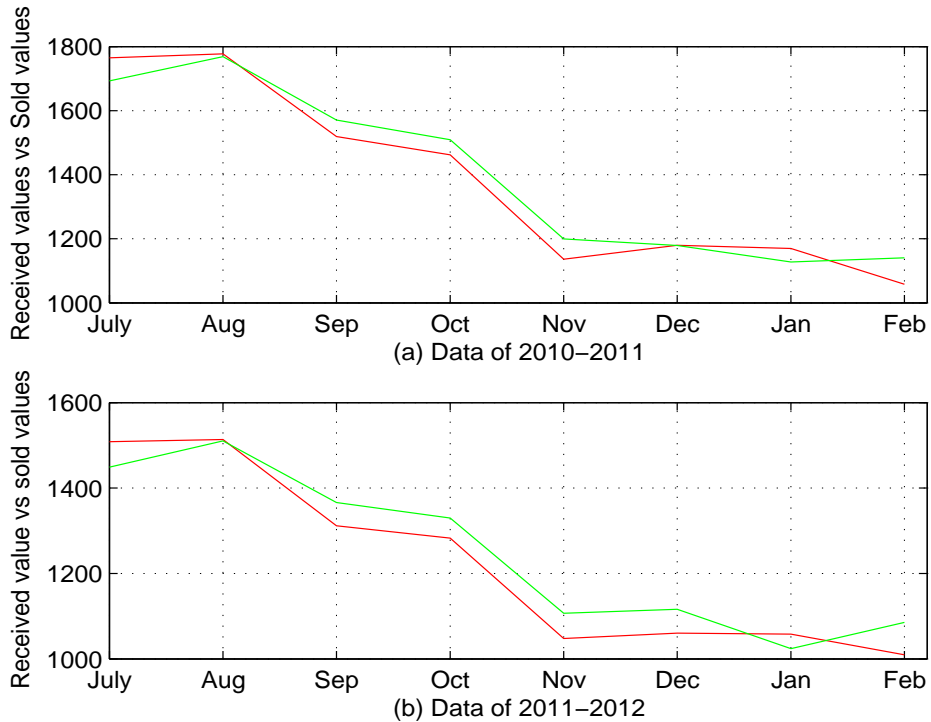}
\caption{Month wise Graphical Representation of losses in a populated city in year 2010-2012.}
\end{figure*}

In [4,5] authors elaborated ways of communication, in how many ways we can transmit the data of smart meter to utility. S. S. S. R Depuru and \textit{et. al.} shown how an electromechanical meter works and why is smart meter better than electromechanical meter.

In [6] central observer meter is placed, which is cost effective because a smart meter is placed at secondary side of transformer. It used matrix based approach in excel to show electricity theft case and normal case. Where as if large amount of data has to be managed than larger matrices will be required. Memory requirements will increase, time consumption to solve large matrices will increase.

[8,9] have some mathematical modeling techniques which helps to detect and control electricity theft using some classifiers. [8] discussed a Graphical User Interface (GUI) based software implemented in Malayesia.

\section{Losses Due To Electricity Theft}

Electricity theft is basically an illegal way of getting the energy for different uses, resulting in loss for utility companies. Losses consist of  technical and non technical losses. There are about \$25 billion of losses annually in the world [1]. Losses can actually be computed by finding the energy supplied, subtracting the amount of energy billed/paid [3]. If we want to calculate non-technical losses (NTL) simply one way of calculating it is to calculate technical losses. We can evaluate it as follows.
\begin{equation}
Total\ Energy\ Losses = Energy\ Supplied - Bills\ paid
\end{equation}
\begin{equation}
Total\ Energy\ Losses = NTL\ +\ TL
\end{equation}
Combining equation 1 and 2, we get
\begin{equation}
 NTL = Energy\ Supplied - Bills\ Paid - TL
\end{equation}
In data $^1$ above shown in table 1. 10.625\% of losses occurred monthly in the year 2011-12 till February [7]. Percentage losses are calculated as:
\begin{equation*}
Percentage\ Loss = \left ( \frac{Received\ Value - Sold\ Value}{Received\ Value} \right)*100
\end{equation*}

Decrease or difference between 2010-2011 and 2011-2012 can be found out by subtracting the value of the present year from the previous year in table 1. and graphically shown in fig 1.
\footnote {Area of a city = 684 sq mile , Population of city = 11,000,000}

There are certain methods of stealing electricity. The core reason of stealing is lack of awareness amongst the peoples, due to which this unpleasant act is being performed in different areas of the world. Meter tempering can be done in electromechanical meters and smart meters as well. Tempering in electromechanical meter is explained in detail below.

\subsection{Theft in Electromechanical Meters}
Few methods of stealing electricity are
\begin{itemize}
\item
Taking connections directly from distribution lines.
\item
Grounding the neutral wire.
\item
Putting a magnet on electromechanical meter like neodymium [1].
\item
Inserting some disc to stop rotating of the coil.
\item
Hitting the meter to damage the rotating coil [2].
\item
Interchanging input output connections.
\end{itemize}

But these disputed issues can be minimized by using the smart meters. Even in smart meter, one can take connections directly from distribution system but smart meters
have the ability to record zero reading. It inform the utility system by sending data through different techniques. These techniques include bluetooth, Power Line Carrier (PLC), Internet protocols. Session Initiation Protocol (SIP) can be used for controlling of Voice over Internet Protocol (VoIP), Zigbee 802.15.4 can be used in Home Area Networks (HANs) [4].

\begin{figure*}[t]
\centering
\includegraphics[height=6cm, width=12cm]{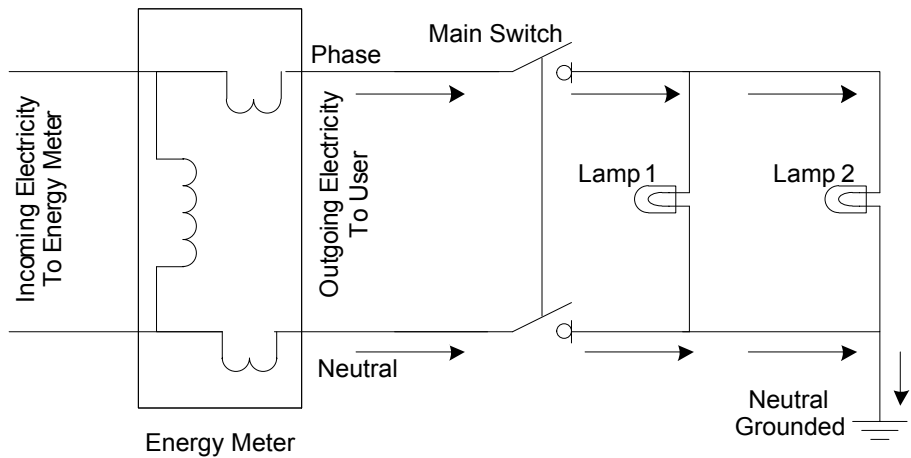}
\caption{Neutral Grounded}
\end{figure*}

In second point if we ground neutral wire then energy meter assume the circuit is not complete and does not measure reading. As we know that moving coil in electromechanical meter can be easily affected by magnetic field lines. So if we put a magnet on electromechanical meter its magnetic field effects the coil motion and cause it to move slow, or even stop if magnet is strong magnet like neodymium. If someone insert an x-ray disc in electromechanical meter, it also interact with coil and affect its performance. Hitting meter shows same results by damaging coil in electromechanical meter. In last point interchanging input output connections in electromechanical meter starts moving in reverse direction, which is also a method to produce less reading, till end of the month.

\subsection{Theft in smart meters}
Smart grid is a very generalized word, it includes diverse kind of sub-infrastructures. One of the important infrastructures is AMI discussed in [2]. Due to many advantages of AMI, every community has the desire to install this system for its ease. AMI is an infrastructure which has many function but it can also be used to control electricity theft. AMI is an infrastructure and smart meter is an entity which can be placed at each and every home/industry, replacing electromechanical Kilo-Watt hour (KWh) meters.

AMI provides a new sensor based approach. If sensors are installed in the electrical equipment, then AMI can be useful in a way that utility or power distributing companies can predict load of a specific area. This is useful for utility in a way that they will design a correct and efficient load flow to certain area. This technique is efficient to save many of economic issues for installing an infrastructure for any area.

Smart meter is a digital device, uses $\mu$-controller and certain other digital instruments. Function of smart meter includes.
\begin{enumerate}
\item
Self billing.
\item
Avoid outages in HAN's.
\item
Remote connect and disconnect.
\item
Remote authentication like sending control messages
\end{enumerate}

While authenticating, data tempering occurs, using software hacking. False authentication can be used to authenticate the password and hack the data from smart meter.

Some hardware hacking, specially designed for fraud purposes are also designed by professionals like descrambler boxes, which reads data from smart meters and are used for illegal purposes.

Time of use capability is also present in AMI, like billing during peak load must be little higher than billing in an off peak load timings.

Methods mentioned for electricity theft in electromechanical meter can be applied to smart meters as well, except putting magnet of neodymium, inserting disc, or hitting it, by this mechanical shock the meter does not work properly.

One of the objections from consumer is that smart meters are used as a spy at our homes. It discloses privacy of our homes, which is not ethically viable. It emits certain kind of radiations which are toxic and dangerous to humans life. It also interferes radio frequency and create problems in radio transmissions to people. Mobile police also uses radio frequency which is interrupted by emission of smart meters [2].

\subsection{Engineered ways of Theft}
 Some of the sophisticated ways of stealing electricity [3] are
 \begin{enumerate}
 \item
 Tempering the current transformers (CT) secondary side of the energy meter,it is generally insulated. where CT's are used to measure current flowing through it. If any one temper CT, then it will not be able to measure correct current passing from energy meter to the consumer or it will record slow readings.
 \item
 Internal calibration of electromechanical energy meter is not correct; the coil used in it is not calibrated correctly.
 \item
 In three phase meters if neutral is kept open, and only one out of three phases is used, than electromechanical meter assume that no energy is flowing through it to the customer. These kind of thefts are easily detected in smart meter by an option of ``EL" glowing.
 \end{enumerate}

EL is an option in smart meter, whose Light Emitting Diode (LED) when flashes shows certain points, such as the miss match between the phase and neutral current is detected by Earth Leakage (EL) LED.
\begin{itemize}
\item
``EL" glows in smart meter means either neutral of your home is connected to the neutral of your neighbors or vice versa.
\item
Phase of your home is connected to the phase of your neighbors or vice versa.
\item
Neutral is connected to the ground.
\end{itemize}

If this ``EL" LED flashes, it will also be visible to the utility, so utility can check the problem manually to control theft.

\section{To Communicate Data To Utility Safely}

Communicating to utility follows a step wise procedure. Smart meter has the ability to measure the energy flowing through it, records the values using micro controller. it updates the values in its registers, but if there occurs any problem in wireless data transfer, it will re-check wireless device. Resolve the issue and and re-update data in smart meter as shown in fig. 2. One of the technical way of theft is to make a meter read slow. If partial electricity is taking by an illegal means, and high energy consumption devices like motors are operated by that electricity. These theft can be examined and checked physically to make all devices operate through legal connection. After that data will be transferred to utility with out letting an intruder to hack it or distort it, through descrambler boxes etc. This data transfer is also an important phase, and it needs attention. Next is to store data at server and make it available for technical computation, that is billing procedure, etc.

\begin{figure}{}
\centering
\includegraphics[height=14cm, width=8cm]{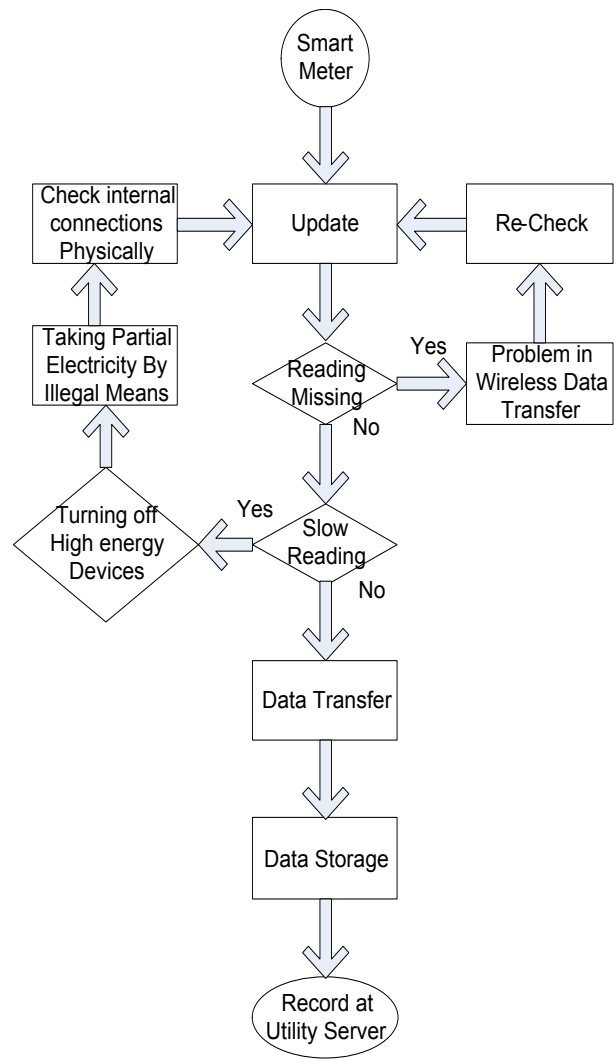}
\caption{Communicating Data To Utility}
\end{figure}

Smart meter has the ability to measure reading time and again, and send it through different techniques like wireless data transfer using different protocols. Bluetooth is one of the method through which we can collect the data from smaller distances such as we can use Bluetooth for HAN as a metering device which follows standard protocols of Bluetooth that is 802.15.1. Broadband Power Line communication (BPL) is another way of communication to the grid, it has certain protocols like Transmission Control Protocol/ Internet Protocol (TCP/IP). It is an advanced form of Power Line Communication (PLC), and it uses a radio frequency spectrum. It causes hurdles in radio communication is one of disadvantage of BPL. Using wired data lines we can also communicate the data from certain industry or home to some central device like smart meter and then send the data wirelessly to the server at utility using Wi-Fi, WiMAX, which follows the standards 802.11g [4].

There are certain other protocols like SIP which supports Voice over Internet Protocol (VoIP), this protocol controls the video and audio data as well. SIP also controls few of the other protocols like Transmission Control Protocol (TCP), Hyper Text Transfer Protocols (HTTP), User Datagram Protocol (UDP). SIP is a very common protocol and dals with many of other protocols. Zigbee is one other protocol which can be used for HAN. It uses standards of 802.15.4. Global System for Mobile communication (GSM), General Packet Radio Server (GPRS) can be another way to send the data to utility using them [5].

By discussing all these ways of communication there will be a problem of huge data transfer through these networks. In wired and wireless services which are using currently IPv4. It uses total of $2^{32}$ addresses, equivalent to 4294967296. These addresses are insufficient to control all devices of all homes of the world including industries etc. On the other hand we have IPv6 which has a total of 128 bits or 32 hex digit code means $2^{128}$ addresses in which there is 48 bit for the node address only, it has a lot of addresses, and we can use them to fulfil requirement of the said scenario. There is one other approach to transmit the data known as Power Line Carrier (PLC). If it is applied through optical fiber then this would be a once and for all investment, and is called as Overhead Power Ground Wire (OPGW).

\begin{figure*}[t]
\centering
\includegraphics[height=12cm, width=16cm]{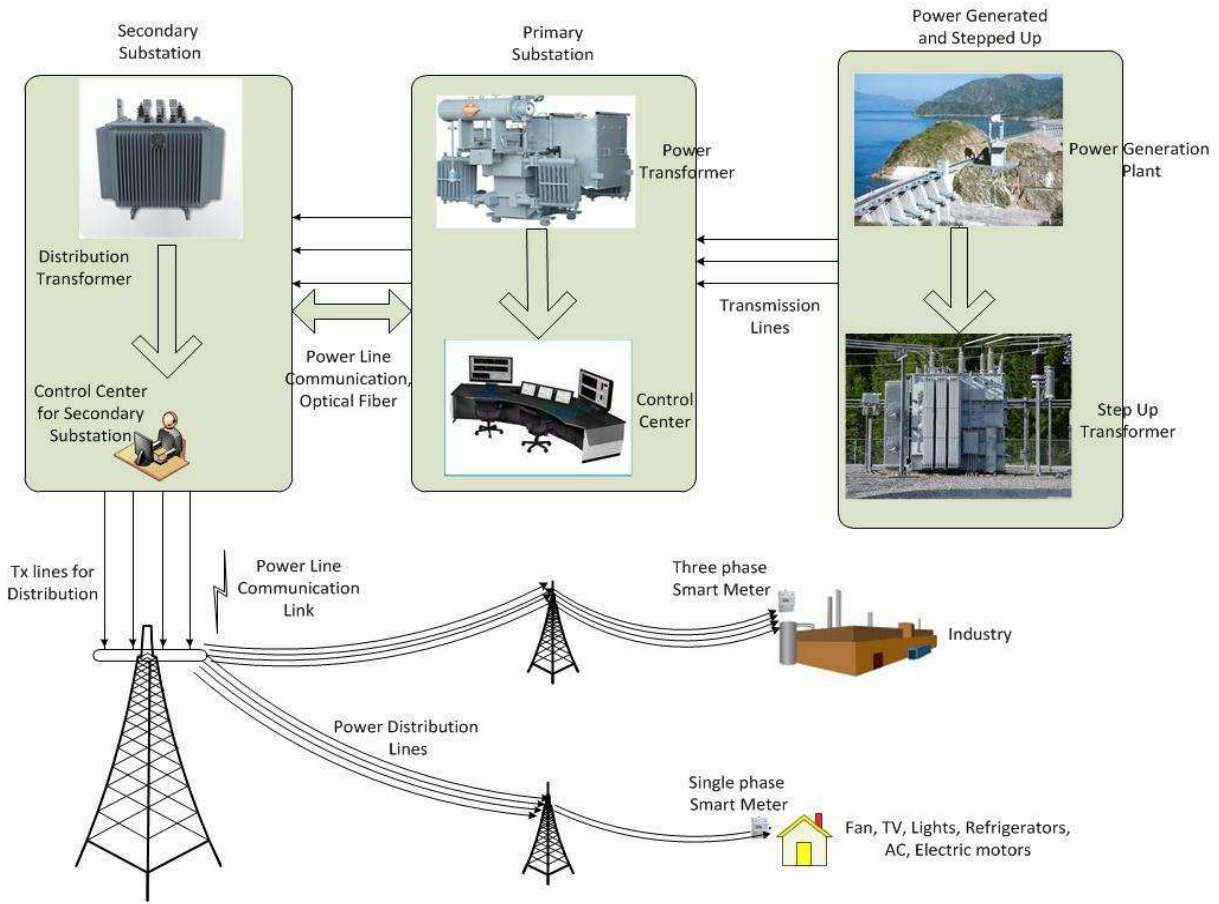}
\caption{Power Flow in Advanced Metering Infrastructure}
\end{figure*}

It would be feasible for controlling the whole of the data if we use one way communication or two way communication from or towards utility. In two way communication we have an advantage of turning on and off the smart meter of any home or industry. Bandim. C. J, \textit{et al.} proposed an idea of low cost methodology of sending the data to the utility, by placing a smart meter on the secondary side of the distribution transformer, which monitors data of home meters locally, records the readings, sends data through wireless device, and communicate two way communication with utility.

\section{Causes And Effects Of Electricity Theft}

Theft is a serious crime, creating short fall, increase of load, decrease of frequency, which is not acceptable and causing load shedding, increase of tariff on the legal customers[1].The main reason behind electricity theft are low literacy rate and lack of awareness. Circular debt is one of the serious consequences of the electricity theft. It can be explained as electricity power is produced through many ways like from oil, turbines are operated and produce electricity. So now if oil is supplied to the utility for their use to run turbines for electricity generation. While utility is not paying to the oil suppliers, and utility is producing electricity from that oil, and selling it to customers. Then losses come into act non paying customers, and theft are also very dangerous, this is what it means that they are not paying the utility back. So this is the issue which is called as circular debt.

In the third world countries, people are mostly not able to pay their utility bills. Government can also help deserving people by avoiding the electricity theft thus providing subsidy to minimize per unit cost. For further improvement in electrical power system respective government needs to give incentives to the capable people to focus on their own electricity production, like emerging technology for production of electricity from solar cells, wind power, hydel etc. Fulfil their own use and sell it to the utility for their own good as well, and be benefited from the Utility by synchronizing their systems successfully.

Power flow mechanism in AMI shown in fig. 4. can briefly be described as power is generated at hydro power plants. As control rooms and control sections are very important part of each and every portion of electricity power generation, primary substation, and secondary substation. Power generation plant is very important part, because whole of electricity is generated at power plants from water. Head is one important issue. Water head and turbine size are mechanical portions. Which also needs to be controlled. These all controls are present for power generation plants on mimic board in control room. Water level is also to be noticed, flow of water through gates. Water reservoirs are kept in water bays for running turbine at peak hours. For control at power generation and control at remote areas Supervisory Control and Data Acquisition System (SCADA)is used. Distributed Control Systems (DCS) is also used for control and supervision at power generation plants. Power flows towards primary substation through transmission lines where it is maintained on the grid in control room.

At Primary substation electricity is stepped down to certain limit. Where electricity could be stolen at any point. At generation less electricity production and waste of water is also a theft. Where system has the ability to produce more electricity, than they were producing. Electricity can also be watched on its way to primary substation. It is quite possible that it could be stolen in a way that is why control centers are deployed and data is observed time to time. After primary distribution it is transmitted to secondary substation, where data comes from the end users, through smart meters or PLC. Home appliances can be controlled using Zigbee.

\section{Mathematical Methods To Control Electricity Theft}

Several methods are used to identify electricity theft using certain mathematical methods like Support Vector Machine LINEAR (SVM-LINEAR), Support Vector Machine- Radial Basis Function (SVM-RBF), Artificial Neural Network- Multi Layer Perceptrons (ANN-MLP), Optimum Path Forest classifier (OPF) [8]. SVM is a regression based technique, in which dependent and independent variables are considered. It defines certain parameters to define a graph or compare it to the standard data or graph. In this method special kind of theft are recognized. If there occurs an abrupt change in load flow it notify that change and store that data as faulty one [7]. Lagrange function can also be the method to study load flow in electrical power system toward distribution. Lagrange basic function can be given as:
\begin{equation}
\pounds = \sum_{i=0}^{N}F_{i}+\lambda \left [ P_{d}-\sum_{i=0}^{N}P_{gi} \right ]
\end{equation}
\begin{equation}
\frac{\mathrm{d} \pounds}{\mathrm{d} P_{gi}}=\frac{\mathrm{d} F_i}{\mathrm{d} P_{gi}}-\lambda
\end{equation}
Taking derivative of equation (4) to solve it for the generation cost of certain operating unit, we get equation (5). It can be the approach for finding technical and non-technical losses in one way as if we consider above equation including losses it will be in a form like:
\begin{equation}
\pounds = \sum_{i=0}^{N}F_{i}+\lambda \left [ P_{d}+P_{L}-\sum_{i=0}^{N}P_{gi} \right ]
\end{equation}
Solving this equation we can find losses which are normally called as technical losses, that is losses on generation side.

Using Lagrange for SVM we can change parameters in this formula for desired situation, further in this method we can get a matrix form, graph of standard form and collect data to identify theft.

Their is one other technique ANN-MLP which is based on modeling techniques, obeys some of non-linear statistical modeling or tree diagram. Other part of it is MLP which is a type of linear classifier and selects better output among outputs from its input.

Ramos, C. C. O \textit {et al.} proposed OPF based technique [8]. It is an approach in which better output is replaced for the previous value selected to reach to identify theft. It needs no parameters to be assumed. Its training phase operation is very fast, an overview is tested by [8] and showed that an OPF has a higher hit rate of theft and having more accuracy than SVM-LINEAR, SVM-RBF, and ANN-MLP [8].

\section{Conclusion}
Electricity thefts are of many types. They are summarized in this paper. Theft can be possible in smart meter as well, which can also be controlled by spreading awareness in peoples on media etc. Some mathematical models are also helpful to detect and control electricity theft.


\begin{thebibliography}{1}

\bibitem{IEEEhowto:kopka}
S. S. S. R Depuru, L. Wang, V. Devabhaktuni. ``Electricity theft: Overview, issues, prevention and a smart meter based approach to control theft."
\hskip 1em plus 0.5em minus 0.4em\relax Energy Policy 39 (2011) 1007-1015.

\bibitem{}
S. McLaughlin, D. Podkuiko, and P. McDaniel. ``Energy theft in Advanced Metering Infrastructure" \hskip 1em plus 0.5em minus 0.4em\relax Pennsylvania State University, University Park.

\bibitem{}
S. S. S. R Depuru, L. Wang, V. Devabhaktuni and N. Gudi. ``Measures and setbacks for controlling electricity theft."

\bibitem{}
S. S. S. R Depuru, L. Wang, V. Devabhaktuni. ``Smart meters for power grid: Challenges, issues, advantages and status." \hskip 1em plus 0.5em minus 0.4em\relax Renewable and sustainable energy reviews 15 (2011) 2736-2742.

\bibitem{}
S. S. S. R Depuru, L. Wang, V. Devabhaktuni and N. Gudi.``Smart meters for power grid: Challenges, issues, advantages and status". \hskip 1em plus 0.5em minus 0.4em\relax IEEE 2011.

\bibitem{}
C. J. Bandim, J. E. R. Alves Jr., A. V. Pinto Jr, F. C. Souza, M. R. B. Loureiro, C. A.Mangalhaes and F. Galvez-Durand. ``Identification of energy theft and tampered meters using a central observer meter: A mathematical approach" \hskip 1em plus 0.5em minus 0.4em\relax IEEE 2003.

\bibitem{}
www.lesco.gov.pk  (12-04-2012)

\bibitem{}
J. Nagi, A. M. Mohammad, K. S. Yap, S. K. Tiog, S. K. Ahmed. ``Non-Technical Loss for Detection of Electricity Theft using Support Vector Machines"
\hskip 1em plus 0.5em minus 0.4em\relax 2nd IEEE international conference on power and energy.

\bibitem{}
C. C. O Ramos, A. N. Souza, J. P. Papa, A. X. Falcao. ``Fast Non-Technical Lasses Identification through Optimum-Path Forest."

\end{thebibliography}
\end{document}